\documentclass[%
aps,
prd,
twocolumn,
superscriptaddress,
nofootinbib,
longbibliography,
floatfix,
amsmath,amssymb,
]{revtex4-2}

\usepackage{graphicx}
\usepackage{dcolumn}
\usepackage{bm}
\usepackage{enumitem}
\usepackage{mathtools}
\usepackage{mathrsfs}  
\usepackage{amsmath}
\usepackage{amssymb}
\usepackage{tensor}
\usepackage{bookmark}
\usepackage[usenames,dvipsnames]{xcolor}
\usepackage{hyperref}
\usepackage{soul}
\usepackage[capitalise]{cleveref}
\usepackage{orcidlink}
\usepackage[normalem]{ulem}

\crefformat{equation}{Eq.~(#2#1#3)}
\crefformat{table}{Tab.~(#2#1#3)}
\crefformat{figure}{Fig.~(#2#1#3)}
\crefformat{appendix}{App.~(#2#1#3)}
\crefformat{section}{Sec.~(#2#1#3)}
\crefformat{subsection}{Subsec.~(#2#1#3)}

\allowdisplaybreaks
\begin{document}

\title{
Nonlinear evolution in Galileon EFTs: Regularization and screening
}

\author{Farid Thaalba}
\affiliation{Nottingham Centre of Gravity \& School of Mathematical Sciences, University of Nottingham, University Park, Nottingham NG7 2RD, United Kingdom}
\affiliation{SISSA, Via Bonomea 265, 34136 Trieste, Italy and INFN Sezione di Trieste}

\author{Aaron~Held\,\orcidlink{0000-0003-2701-9361}}
\affiliation{
Institut de Physique Théorique Philippe Meyer, Laboratoire de Physique de l’\'Ecole normale sup\'erieure (ENS), Universit\'e PSL, CNRS, Sorbonne Universit\'e, Universit\'e Paris Cité, F-75005 Paris, France
}

\author{Ramiro Cayuso}
\affiliation{SISSA, Via Bonomea 265, 34136 Trieste, Italy and INFN Sezione di Trieste}
\affiliation{IFPU - Institute for Fundamental Physics of the Universe, Via Beirut 2, 34014 Trieste, Italy}

\author{Thomas P. Sotiriou}
\affiliation{Nottingham Centre of Gravity \& School of Mathematical Sciences, University of Nottingham, University Park, Nottingham NG7 2RD, United Kingdom}
\affiliation{School of Physics and Astronomy, University of Nottingham, University Park, Nottingham NG7 2RD, United Kingdom}

\begin{abstract}
In a scalar effective field theory (EFT) with Galileon symmetry, we distinguish the strong-field expansion from the derivative expansion and study spherically symmetric classical nonlinear dynamics. More concretely, we consider two models related via perturbative field redefinitions: one with second-order equations that can also exhibit screening, and one that includes a higher-derivative term that introduces a propagating ghost. The latter term is of the type that can ``regularise'' the equations to render them well-posed as an initial value problem. For initial data that respect the derivative expansion, we find that the first model (ghost-free, unregularised) develops ill-posed regions during evolution only when the derivative expansion breaks down. This result persists in the strong-field regime. We further find that the regularized model reproduces the evolution of the unregularised one when the latter remains well-posed, while it exhibits tachyonic behaviour when the unregularised one becomes ill-posed. We also consider initial data that corresponds to stationary states that exhibit screening in the unregularised, ghost-free theory. In this case, we find that the regularised theory can remain well-posed for initial data that the unregularised one developed ill-posed regions. However, the ghost term naturally dominates over the Galileon term in the screening regime, contrary to common expectation.

\end{abstract}

\maketitle

\section{Introduction}
\label{sec:intro}

The growing catalogue of gravitational-wave (GW) events obtained by the LIGO-Virgo-KAGRA network~\cite{LIGOScientific:2018mvr,LIGOScientific:2020ibl,KAGRA:2021vkt} has opened the dynamical strong-field regime of gravity to direct observation, and the next generation of detectors~\cite{LISA:2024hlh,ET:2025xjr} will considerably extend this dataset, both in quality and quantity.
The most informative signals are produced precisely where the field is strong, and the dynamics are nonlinear. A maximally informative comparison between theory and observation thus requires robust nonlinear strong-field predictions, both in general relativity (GR) and --- if we seek to constrain potential deviations --- also when corrections are included.
In practice, this requires robust numerical relativity (NR) solutions of the full nonlinear equations~\cite{LIGOScientific:2021sio}.

Effective field theory (EFT) provides a systematic framework to account for these corrections~\cite{Donoghue:1994dn,Burgess:2003jk}. Once the low-energy field content and symmetries are fixed, the EFT is maximally agnostic. Denoting an arbitrary set of low-energy matter fields by $\Phi$, a general covariant EFT (with or without dynamical metric $g$)\footnote{
    We work in mostly plus signature and follow standard notation for the associated Riemann tensor $R_{abcd}$, Weyl tensor $C_{abcd}$, Ricci tensor $R_{ab}$, and Ricci scalar $R$.
}
may be written schematically as 
$\mathcal{L}=\mathcal{L}_{0}+\sum_{k>2}^{N}\Lambda^{4-k}\sum_i c_{k,i}\mathcal{O}_{k,i}$,
where $\mathcal{L}_{0}[g,\Phi]$ contains the leading dynamics\footnote{
    For instance, for a single scalar field $\Phi=\{\phi\}$, the leading two-derivative Lagrangian $\mathcal{L}_{0}^{(\phi)}\supset\frac{1}{2}(\partial\phi)^2$ starts at order $k=4$ and for gravity we have $\mathcal{L}_{0}^\text{(gravity)}= M_{\text{Pl}}^2 R$.
}, while the operators $\mathcal{O}_{k,i}[g,\Phi]$ form a basis of all local covariant scalars  of dimension $k$. The dimensionless Wilson coefficients $c_{k,i}$ encode the ultraviolet completion, and the scale $\Lambda$ controls the EFT expansion. For simplicity, we have assumed a single-scale, i.e., a mixed derivative/field/curvature expansion\footnote{
    One may alternatively distinguish between curvature and derivatives, i.e., $\mathcal{L}{=}M_{\text{Pl}}^2R{+}\sum_{m \geq 2,n \geq 0}\Lambda^{4-2m}_{R}\Lambda_\partial^{-n} \mathcal{L}_{(m,n)}$, where the scales $\ell_R$ and $\ell_\partial$ now separately control an expansion in curvature from an expansion in external derivatives and $\mathcal{L}_{(m,n)}$ schematically denotes a basis of all possible covariant scalars at order $(m,n)$. Identifying $\Lambda{=}\Lambda_R{=}\Lambda_\partial$ and $k{=}2m{+}n$ recovers the joint expansion.
    Similarly, one may distinguish powers of the respective fields~$\Phi$.
} but, as we will spell out in more detail, in some physical situations it may become important to distinguish these expansions.

Faithfully describing dynamics with an EFT, at some finite truncation order, requires formulating its equations as an initial value problem (IVP). For the evolution to be predictive, this IVP needs to be well-posed~\cite{hadamard}. Generally, one faces two complications. The first one is that the number of degrees of freedom does not necessarily match the explicit EFT field content, as the higher order derivative terms introduce additional propagating degrees of freedom. At a given canonical dimension $k$, the differential order, and hence the total number of degrees of freedom, is controlled by the operators with the largest number of external derivatives, i.e., $\Phi\Box^{k/2-1}\Phi$ for matter as well as $R\Box^{k/2-2}R$ and $R_{ab}\Box^{k/2-2}R^{ab}$ for gravity. These additional degrees of freedom are usually considered spurious, as artefacts of the truncation, and their dynamics are assumed to compromise well-posedness \footnote{The loss of well-posedness is often attributed to the fact that the additional degrees of freedom have opposite-sign kinetic terms (so-called ghosts), thereby yielding an unbounded Hamiltonian, which is then interpreted as a definitive sign of an instabilities~\cite{Ostrogradsky:1850fid,Woodard:2015zca}. Note, however, that there are counterexamples to the latter~\cite{Deffayet:2021nnt,Deffayet:2023wdg,Deffayet:2025lnj,Held:2025fii,Deffayet:2026cnu,Deffayet:2026uoe,Ewasiuk:2026kmz}.}. Subclasses of EFTs that maintain second-order equations do exist. These include any non-derivative self-interactions and subsets of derivative interactions that have been mapped out extensively~\cite{Horndeski:1974wa, Kobayashi:2019hrl, Deffayet:2009mn, Lovelock:1971yv, Padmanabhan:2013xyr}. However, even derivative interactions within this subset compromise quasilinearity --- a key assumption of standard well-posedness proofs --- and make it hard to formulate a well-posed IVP. This is the second complication.

The distinction between the two types of EFT terms discussed above, those that introduce spurious degrees of freedom --- henceforth referred to as ghostly for brevity --- and those that do not, is important for phenomenology. The non-ghostly terms are known to lead to interesting nonlinear phenomena, such as  Vainshtein screening~\cite{Vainshtein:1972sx,Deffayet:2001uk,Babichev:2013usa} and spontaneous scalarisation~\cite{Silva:2017uqg, Doneva:2017bvd, Doneva:2022ewd}, assuming that the ghostly terms can remain subdominant. We remark however that there is in general no symmetry principle that separates the two types of terms. For example, consider $(\Box \phi)^2$ and the cubic Galileon term $\left(\partial\phi\right)^2\!\left(\Box\phi\right)$ in a scalar theory (similarly for $R^3$ and $R\Box R$ in pure gravity).

The narrative above treats the degrees of freedom introduced by the ghostly terms as intrinsically problematic. Indeed, most attempts to address the well-posedness issue in EFTs, such as perturbative order reduction~\cite{Witek:2018dmd,Okounkova:2019dfo}, dissipative schemes based on higher-order Lorentz-violating spatial derivatives~\cite{deRham:2023ngf}, and the ``fixing-the-equations'' methods~\cite{Cayuso:2017iqc,Cayuso:2020lca}, aim to remove these degrees of freedom or heavily suppress their effect on the dynamics.  It has, however, recently been proven in the context of pure gravity that ghostly terms can systematically be added to EFTs to render them well-posed \cite{Figueras:2024bba}. The specific terms are referred to as  ``regularisation" terms, and they can be understood to arise via suitable perturbative field redefinitions that are expected to preserve observables and map solutions onto each other, order by order in perturbation theory. Finding regularization terms for nonlinear scalar field theories is straightforward as well; see, e.g.,~\cite{Held:2025fii} for a review.

An interesting question is what happens if one attempts to apply this approach to a theory and physical scenario in which some of the non-ghostly EFT terms are expected to be dominant? Scenarios in which low-order EFT corrections become significant and lead to large nonlinearities arguably make the most interesting EFTs phenomenologically, as discussed above.  Can well-posed evolution be established? And if yes, do the regularizations/ghostly terms remain subdominant as expected in such scenarios? This last question relates to the central question of whether field-redefinition equivalence remains intact once strong-field phenomena become relevant. 
Exploring these questions is the main goal of this study.

A  particularly clean setting in which to address all of the above questions, and also get a clear exposion of the workings of the regularization approach, is the EFT for a single scalar field on a fixed flat background in which the scalar field is assumed to be invariant under Galileon symmetry~\cite{Nicolis:2008in}, i.e., $\phi\mapsto\phi{+}b_\mu x^\mu{+}c$. Here, the leading corrections to both operator sectors are cleanly separated. The cubic Galileon interaction $\left(\partial\phi\right)^2\!\left(\Box\phi\right)$ generates nonlinear derivative interactions while maintaining second-order field equations. In contrast, the leading regularisation operator $\phi\Box^2\phi$ introduces an additional fiducial mode but does not contribute to nonlinear interactions in flat space. 
Moreover, the ghost-free Galileon equations are known to exhibit ill-posed regions in the sufficiently nonlinear regime~\cite{Bernard:2019fjb,Gerhardinger:2022bcw}. The model thus isolates the relation between nonlinear hyperbolicity, higher-derivative regularisation, and perturbative field-redefinition equivalence without the additional difficulties introduced by dynamical gravity. We comment on extensions to, and expectations for, scalar-tensor theory and pure gravity in the conclusions.

At the same time, the Galileon is not merely a proxy for the gravitational problem. Its derivative interactions support Vainshtein-screened configurations and thus provide a direct dynamical test of the hierarchy on which screening is conventionally based~\cite{Joyce:2014kja}. 

In the following, we solve the spherically symmetric nonlinear dynamics on both sides of the perturbative field redefinition: in a lower-derivative ghost-free formulation and in a higher-derivative ghostly formulation. This allows us to determine when the two descriptions remain physically equivalent, how their behaviour depends separately on the field and derivative expansion, and whether the regularisation terms can remain parametrically subdominant inside a screened region.

\section{Setup: A scalar field with Galileon symmetry}
\label{sec:setup}

We focus on a single scalar field $\phi$ in four-dimensional Minkowski space, invariant under the Galileon symmetry $\phi\mapsto\phi{+}b_\mu x^\mu{+}c$. Up to total derivatives, the leading terms compatible with this symmetry are the standard kinetic term, the higher-derivative ``ghost'' term, and the cubic Galileon interaction,
\begin{align}
    \mathcal{L} &=
    -\frac{1}{2}\left(\partial\phi\right)^2
    -\beta\left(\Box\phi\right)^2
    +\alpha\left(\partial\phi\right)^2\!\left(\Box\phi\right)
    \;,
    \label{eq:Lagrangian-Galileon-at-face-value}
\end{align}
where $\alpha$ and $\beta$ are dimensionful coupling constants of canonical mass dimension $[\alpha]=3$ and $[\beta]=2$. We chose to measure all other dimensionful quantities in units of $\alpha^{1/3}$ and, in practice, perform all numerical simulations with $\alpha=1$ and the respective dimensionless ratios for all other parameters. Having thereby used $\alpha$ to set the scale, the theory space is fully determined by a single dimensionless ratio $\beta/\alpha^{2/3}$. We refer to $\beta/\alpha^{2/3}{=}0$ as the unrgularised and to any $\beta/\alpha^{2/3}{\neq}0$ as the regularised theory.

As our motivation is twofold, we seek to understand the respective dynamical evolution problem, both as a purely technical probe of regularisation and in the context of the EFT. We detail these two points of view in~\cref{sec:regularisation} and~\cref{sec:EFT}, respectively, and then discuss suitable classes of initial data in~\cref{sec:ID}.

We will also consider a conformally coupled source Lagrangian $\mathcal{L}_\text{source}{=}\gamma\phi T$, where $T$ is the trace of an energy-momentum tensor that we consider to be static and $\gamma$ denotes the respective conformal coupling. All details concerning this source term are discussed in~\cref{sec:screening-ID}.

\subsection{Regularising ill-posed regions: a proxy for the gravitational problem}
\label{sec:regularisation}

In the first reading, we take~\eqref{eq:Lagrangian-Galileon-at-face-value} at face value and consider the impact that the regularisation term has on hyperbolic time evolution.
The cubic Galileon alone ($\beta{=}0$) has second-order field equations, but, as we recall in~\cref{sec:unregularised}, these can lose hyperbolicity during evolution, so that the classical time evolution can cease to exist for perfectly regular initial data. The higher-derivative term $\beta(\Box\phi)^2$ is then introduced by hand as a \emph{regularising} operator: it raises the order of the equations, and, as we show below, renders the local IVP well-posed at the price of a fiducial ghost. In this reading, $\beta$ is a control parameter, and the model is a flat-space proxy for the regularisation programme pursued for gravitational EFTs~\cite{Figueras:2024bba,Figueras:2025wtx}, where the same trade-off between well-posedness and additional degrees of freedom appears. In contrast to the scalar-field example at hand, regularisation in gravitational problems is entangled with the technical complications of a dynamical metric. Everything we find in this simpler arena is meant to inform that harder problem.

We emphasize that it is possible to define how far the regularisation is a successful way to deal with ill-posedness, independent of the physical interpretation. Physical interpretation, for instance, in the context of an EFT (see~\cref{sec:EFT}) is a subsequent question.
Operationally, we probe two aspects of regularisation: 
\begin{itemize}
    \item First, we check if the regularisation is necessary, i.e., if the unregularised dynamics fails due to ill-posedness, either on the given set of initial data, or at some later stage during the nonlinear evolution.
    \item Second, we consider the regularisation prescription to be successful (or at least partially successful) if the resulting nonlinear solution becomes independent of $\beta$ in the limit of small regularisation parameter $\beta$ and persists for all future times (or at least for longer than the unregularised case). In practice we probe towards the limit of $\beta\to 0$ and ask that $\beta$ drops below any other scale of interest.
\end{itemize}
In the remainder of this section, we review the initial value formulations of the unregularised ($\beta=0$) and regularised ($\beta\neq0$) dynamics.

\subsubsection{The unregularised dynamics}
\label{sec:unregularised}

In the unregularised case ($\beta=0$), the field equations are second order,
\begin{align}
    \Box \phi {+} \alpha\left(
        (\Box \phi)^2
        {-}
        \partial_{a}\partial_{b}\phi\,\partial^{a}\partial^{b}\phi
    \right)
    = 0
    \;.
    \label{eq:field-eqs_unregularised}
\end{align}
It is well known~\cite{Bernard:2019fjb} that local well-posedness of~\eqref{eq:field-eqs_unregularised} depends on the local field values, so that ill-posed regions can develop during time evolution. The time evolution can therefore cease to exist dynamically. To see this, consider linearising in a perturbation $\delta\phi$ and keeping only the highest-derivative terms, the perturbation obeys $P^{ab}\,\partial_a\partial_b\,\delta\phi+\dots=0$, with the effective inverse metric
\begin{align}
    P^{ab} = \left(1+2\alpha\Box\phi\right)\eta^{ab} - 2\alpha\,\partial^a\partial^b\phi\;.
    \label{eq:effective-metric}
\end{align}
The characteristics are the covectors $\xi_a$ that solve $P^{ab}\xi_a\xi_b=0$. In spherical symmetry, writing the characteristic speed as $c{\coloneqq}{-}\xi_t/\xi_r$, this quadratic reads $P^{tt}c^2~-~2P^{tr}c~+~P^{rr}~=~0$, whose roots, once the spherical reduction of $P^{ab}$ is carried out, are the speeds~\cite{Kreiss,Sarbach:2012pr}
\begin{align}
    c_{\pm} &=
    \frac{1}{r + 2 \alpha r\,\Delta\phi}
    \Big(
        -2 \alpha r\,\partial^2_{tr}\phi
        \notag\\* &\quad
        \pm \sqrt{
            \left(r + 2 \alpha\,\partial_r \phi\right)^2
            + 8 \alpha^2 (\partial_{r}\phi)^2
        }
    \Big)
    \;,
    \label{eq:characteristic-speeds}
\end{align}
where $\Delta$ denotes the spatial Laplacian.
Two features of~\eqref{eq:effective-metric} are worth isolating. First, the coefficient of the second time derivative is $P^{tt}~=~-~(1~+~2\alpha\Box\phi)~-~2\alpha\,\partial_t^2\phi$. For a static or slowly varying background, it vanishes when $1+2\alpha\Box\phi{=}0$, which is exactly where the denominator of~\eqref{eq:characteristic-speeds} vanishes and the speeds $c_\pm$ diverge, i.e., a Keldysh-type transition occurs. 
Second, the same condition $1+2\alpha\Box\phi{=}0$ is also where the effective ghost mass (derived in~\cref{sec:regularised},~\cref{eq:eff-masses}) passes through zero. The loss of hyperbolicity in the unregularised theory and the onset of the tachyon in the regularised theory therefore appear to be related. They may describe the same physical breakdown, as observed in the two respective formulations.

\subsubsection{The regularised dynamics}
\label{sec:regularised}

In the regularised theory, the field equations are fourth order in derivatives. Introducing the fiducial scalar $\psi{\coloneqq}\Box\phi$ and the fiducial one-form $A_{a}{\coloneqq}\partial_a\phi$, they can be written as the second-order system
\begin{align}
    \Box \psi &=
    \frac{1}{\beta}\,\psi
    + \frac{\alpha}{\beta}\left(
        \psi^2
        - \partial_{a}A_{b}\,\partial^{a}A^{b}
    \right)
    \;,
    \notag\\*
    \Box \phi &= \psi
    \;,
    \quad\quad\text{with}\quad
    \Box A_{a} = \partial_{a}\psi
    \;,
    \label{eq:field-eqs_regularised}
\end{align}
where $1/\beta$ now appears as a mass term for $\psi$. In contrast to the unregularised case, the PDE system~\eqref{eq:field-eqs_regularised} is locally well posed around all sufficiently smooth initial data since it constitutes a system of coupled nonlinear wave equations~\cite{1952AcMa...88..141F}. 

As we will see, the ill-posedness of the unregularised theory corresponds to the onset of a linear tachyonic instability in the regularised theory. To make this explicit, consider linear perturbations
$\delta \Phi {\coloneqq} \left(\delta \phi,\delta \psi,\delta A_{a}\right)^T$ of~\eqref{eq:field-eqs_regularised} around a background $\Phi^{\text{BG}}{\coloneqq}(\phi^{\text{BG}},\psi^{\text{BG}},A^{\text{BG}}_a)^T$. Retaining the mass-like (non-derivative) couplings, we find $\Box\,\delta\Phi-\mathcal{M}\,\delta\Phi+\dots{=}0$, where the ellipsis stands for higher-order perturbations and terms involving derivatives of the background, and
\begin{align}
    \mathcal{M} = \begin{pmatrix}
        0 & 1 & 0\\
        0 & \left(1+2 \alpha \psi^{\text{BG}}\right)/\beta~ & 0 \\
        0 & 0 & 0
    \end{pmatrix}~.
    \label{eq:mass-matrix}
\end{align}
The eigenvalues of $\mathcal{M}$ are the effective squared masses of the perturbations,
\begin{align}
    m^2_{\phi} = 0~, \quad m^2_{\psi} = \frac{1+2 \alpha \psi^{\text{BG}}}{\beta}~, \quad  m^2_{A} = 0~.
    \label{eq:eff-masses}
\end{align}
Aside from the two massless modes, the ghost mode $\psi$ has a background-dependent effective mass. To avoid a tachyon around the trivial vacuum ($\psi^{\text{BG}}{=}0$) we need $\beta{>}0$.

\subsection{Galileons as effective field theories}
\label{sec:EFT}

In the second reading,~\eqref{eq:Lagrangian-Galileon-at-face-value} is the leading part of a genuine EFT for a scalar with Galileon symmetry~\cite{Nicolis:2008in,Deffayet:2011gz,Joyce:2014kja}. Now the higher-derivative term is not a regulator but simply the leading operator in the derivative expansion, expected with a dimensionless coefficient of order unity. Whenever this term is negligible, the field content is fixed: a single, healthy scalar with second-order dynamics. The cubic Galileon interaction then produces Vainshtein screening~\cite{Vainshtein:1972sx,Deffayet:2001uk,Babichev:2013usa}, the mechanism by which the nonlinearities suppress the scalar force near a source and restore agreement with local tests. The standard treatment of screening assumes precisely that the ghost/higher-derivative operators can be discarded around the screened solution~\cite{Nicolis:2004qq,Luty:2003vm,Joyce:2014kja}.

To interpret our results in this context, we can restate the Lagrangian in~\cref{eq:Lagrangian-Galileon-at-face-value} as an EFT, i.e.,
\begin{align}
    \mathcal{L} &=
    -\frac{1}{2}\left(\partial\phi\right)^2
    -\frac{c_\beta}{\Lambda_\partial^2}\left(\Box\phi\right)^2
    +\frac{c_{\alpha}}{\Lambda_\partial^2\Lambda_\phi}\left(\partial\phi\right)^2\!\left(\Box\phi\right)
    \notag\\*[-0.5em]&\quad
    +\mathcal{O}\!\left(\partial^4,\phi^4\right)\!,
    \label{eq:Lagrangian-EFT-Galileon}
\end{align}
where $\mathcal{O}\!\left(\partial^4,\phi^4\right)$ denotes further terms with at least four derivatives or at least four fields, including, for instance, the quartic Galileon. The two scales $\Lambda_\partial$ and $\Lambda_\phi$ control, respectively, a derivative expansion and a field expansion; see \cref{sec:multi-scale-EFT} for further discussion of this distinction. The dimensionless EFT coefficients $(c_\alpha,c_\beta)$ are related to the dimensionful couplings $(\alpha,\beta)$ by multiplying with appropriate powers of the two EFT scales $(\Lambda_\partial,\Lambda_\phi)$, i.e.,  $c_\alpha{=}\alpha\Lambda_\partial^2\Lambda_\phi$ and $c_\beta{=}\beta\Lambda_\partial^2$ are written in terms of the dimensionful parameters $\alpha$ and $\beta$. Under the assumption of order-one EFT coefficients, i.e., $c_i\sim1$ for $i={\alpha,\beta}$, and since we only retain the two leading terms, each pair of $(\alpha,\beta)$ equivalently specifies $(\Lambda_\partial,\Lambda_\phi)$.
%
%
%

\subsubsection{Distinguishing derivative and field expansion}
\label{sec:multi-scale-EFT}

We distinguish between a derivative expansion (labelled by the number of derivatives $n$) and a field expansion (labelled by the number of fields $m$). The ghost term is of order $(n,m){=}(4,2)$, while the Galileon term is of order $(n,m){=}(4,3)$. The two expansions are controlled by the scales $\Lambda_\partial$ and $\Lambda_\phi$, respectively. We deliberately maintain this two-scale expansion, since we want to analyze how these two directions in power counting are probed differently by the dynamics.

If one collapses the double expansion into a single power-counting expansion, the ghost and Galileon terms are of order $n{+}m{=}6$ and $n{+}m{=}7$, respectively.

\subsubsection{Field redefinitions}
\label{sec:field-redefinitions}

The ghost term can be modified by a perturbative \emph{linear} field redefinition
\begin{align}
    \phi \mapsto \phi - \frac{c_\text{redef}}{\Lambda_\partial^2}\,\Box\phi\;,
    \label{eq:field-redef}
\end{align}
under which the kinetic term generates a contribution that shifts $\beta{\mapsto}\hat{\beta}$ (equivalently $c_\beta{\mapsto}\hat{c}_\beta$). In particular, for $c_\text{redef}{=}{-}c_\beta$ the ghost term is removed, $\hat{\beta}{=}0$ (equivalently $\hat{c}_\beta{=}0$). All other terms generated by~\eqref{eq:field-redef} are at least of order $(n,m){=}(6,2)$, hence higher order in derivatives~\footnote{
    The Galileon interaction can be removed by a field redefinition as well, that is, by $\phi\,{\mapsto}\,\phi{-}c_{\alpha}/(\Lambda_\partial^2\Lambda_\phi)(\partial\phi)^2$, under which the generated remainders are at least of order $(n,m){=}(6,3)$. However, while the ghost term can be removed with a \emph{linear} field redefinition, removing the Galileon term requires a \emph{nonlinear} one.
}, and can be absorbed in $\mathcal{O}\!\left(\partial^4,\phi^4\right)$.
From an EFT perspective, $\hat{\beta}{=}0$ is an unnaturally small value, since no symmetry protects it. The unregularised theory should therefore only be trusted as long as the field redefinition in~\eqref{eq:field-redef} remains perturbatively controlled.

In perturbation theory, one expects that linear field redefinitions leave invariant the physics of small perturbations around the leading-order vacuum, here the trivial vacuum $\phi{=}0$. Below we recover this expectation in the full nonlinear time evolution (see also~\cite{Deffayet:2025lnj,Figueras:2025wtx,Held:2025fii}), and we quantify how the equivalence breaks down as nonlinearities in the fields and the derivatives become important. In particular, whenever $|c_\text{redef}\Lambda_\partial^{-2}\Box\phi|{\gtrsim}|\phi|$, the field redefinition in~\eqref{eq:field-redef} need no longer leave the physics invariant.

\subsubsection{EFT diagnostics}
\label{sec:diagnostics}
To make ``perturbatively controlled'' quantitative, we track the local size of the individual interaction terms in the Lagrangian. Rendered dimensionless by dividing by $\Lambda_\partial^n\Lambda_\phi^m$ in accordance with the $(n,m)$ counting, these are
\begin{align}
    \varepsilon_\partial(t,\vec{x})
    &\equiv
    \frac{|(\Box\phi)^2|}{\Lambda_\partial^4\Lambda_\phi^2}
    \;,
    \notag\\
    \varepsilon_{\rm Gal}(t,\vec{x})
    &\equiv
    \frac{|(\partial\phi)^2(\Box\phi)|}{\Lambda_\partial^4\Lambda_\phi^3}
    \;.
    \label{eq:diagnostics}
\end{align}
The quantity $\varepsilon_\partial$ measures the local strength of the derivative expansion and $\varepsilon_{\rm Gal}$ that of the field expansion. We stress that $\sqrt{\varepsilon_\partial}$ also controls the EFT consistency of the field redefinition in~\eqref{eq:field-redef}: it is precisely the ratio $|c_\text{redef}\Lambda_\partial^{-2}\Box\phi|/|\phi|$ that governs whether removing the ghost is a legitimate operation. Tracking $\varepsilon_i$ (with $i{=}\{\partial,\text{Gal}\}$) therefore lets us locate, locally, the regions in which the EFT is extrapolated beyond its cutoff, and separately diagnose which of the two expansions is responsible. Alternatively, instead of~\eqref{eq:diagnostics}, we may construct the diagnostic parameters as ratios of the interaction terms in~\eqref{eq:Lagrangian-EFT-Galileon} to the kinetic term. However, we avoid using these ratios because they diverge when the kinetic term evaluates to zero on the solution; thus, they can diverge without necessarily indicating the breakdown of the EFT.

\subsubsection{Degrees of freedom and energy budget}
\label{sec:dof}
The higher-derivative term in~\eqref{eq:Lagrangian-EFT-Galileon} carries a hidden degree of freedom. A Lagrangian that depends nondegenerately on second time derivatives propagates, by Ostrogradsky's theorem~\cite{Ostrogradsky:1850fid,Woodard:2015zca}, an extra mode whose Hamiltonian is unbounded from below, that is, a ghost. To make this degree of freedom explicit, we introduce an auxiliary field $\chi$ and write
\begin{align}
    \label{eq:L_aux_2}
    \mathcal{L}_{\text{aux.}} &= X + \alpha X \Box \phi - \chi \Box \phi + \frac{1}{2\beta}\chi^2~,
\end{align}
with $X{\coloneqq}{-}\tfrac{1}{2}(\partial\phi)^2$. Integrating $\chi$ out through its algebraic equation of motion $\chi{=}\beta\Box\phi$ returns the ghost term in~\eqref{eq:Lagrangian-EFT-Galileon}, so~\eqref{eq:L_aux_2} is classically equivalent to the original theory for $\beta{\neq}0$. The kinetic matrix of $(\phi,\chi)$ is off-diagonal; the field redefinition
$\phi\,{\to}\,\phi_2{-}\phi_1$ and $\chi\,{\to}\,\phi_2$ diagonalises it,
\begin{align}
    \label{eq:L_diag}
    \mathcal{L} &= X_1 - X_2 + \frac{1}{2\beta}\phi_2^2 + \alpha X \Box \phi~,
\end{align}
where $X_i$ is the kinetic term of $\phi_i$. As expected, there are two degrees of freedom with opposite-sign kinetic terms: the healthy mode $\phi_1$ and the ghost $\phi_2$. The term $\phi_2^2/(2\beta)$ is a mass term for the ghost, with mass squared of order $1/\beta$. As $\beta\to0^+$, the ghost becomes infinitely heavy and, provided nothing excites it, decouples from the low-energy dynamics; this is the field-theoretic content of removing the higher-derivative term by the redefinition~\eqref{eq:field-redef}.

Because it will be informative to follow where energy resides during the evolution, we define the energy $E{\coloneqq}\int \mathrm{d}^3x\, T_{00}$, with $T_{ab}$ the canonical energy-momentum tensor of~\eqref{eq:L_diag} 
\begin{align}
    T_{ab} = T^{\phi_1}_{ab} + T^{\phi_2}_{ab} + T^{\text{int.}}_{ab}~,
\end{align}
where, 
\begin{align}
    T^{\phi_1}_{ab} = \partial_{a}\phi_1\partial_{b}\phi_1-\frac{1}{2}\eta_{ab}\partial_{c}\phi_1\partial^{c}\phi_1~, 
\end{align}
\begin{align}
    T^{\phi_2}_{ab} = -\partial_{a}\phi_2\partial_{b}\phi_2+\frac{1}{2}\eta_{ab}\partial_{c}\phi_2\partial^{c}\phi_2 -\frac{1}{2 \beta}\eta_{ab}(\phi_2)^2~,
\end{align}
and the ``interaction'' piece is
\begin{align}
    T^{\text{int.}}_{ab} &= \alpha \left(\partial_{a}\phi\partial_{b}\phi\Box\phi - \partial_{b}\phi\partial_{c}\partial_{a}\phi\partial^{c}\phi \right. \nonumber \\
    & \left.- \partial_{a}\phi\partial_{c}\partial_{b}\phi\partial^{c}\phi+\eta_{ab}\partial^{c}\phi\partial_{d}\partial_{c}\phi\partial^{d}\phi\right)~.
\end{align}
We split the total energy of the system into three components as 
\begin{align}
    \label{eq:energy-budget}
    E_{\text{tot}} = E_{\phi_1} + E_{\phi_2} + E_{\text{int}}~,
\end{align}
where 
\begin{align}
    E_{\phi_1} &\equiv \int \mathrm{d}^3x T^{\phi_1}_{00}~, \\
    E_{\phi_2} &\equiv \int \mathrm{d}^3x T^{\phi_2}_{00}~, \\
    E_{\text{int.}} &\equiv \int \mathrm{d}^3x T^{\text{int.}}_{00}~.
\end{align}
The ghost contributes with the opposite sign, so a growing $|E_{\phi_2}|$ compensated by $E_{\phi_1}$ is the energetic signature of the instability we discuss below.

\subsection{Spherically-symmetric initial data}
\label{sec:ID}

We probe the two questions above by obtaining finite-time numerical solutions to the corresponding spherically symmetric initial data problems. Given the nature of numerical methods, we do not obtain a rigorous statement that holds for all future times and all initial data. Rather, we choose a sufficiently representative set of initial data families that we detail below. Naturally, our statements are also confined to spherical symmetry. 

Each family of initial data is characterised by a set of parameters, which we normalise by multiplying with appropriate powers of $\alpha$, such that the parameter space of the respective problem is described by a set of dimensionless ratios detailed below. We recall that throughout, the theory space is fully characterised by a single dimensionless ratio $\tilde{\beta}=\beta/\alpha^{2/3}$.

For each simulation, we evaluate the diagnostics $\varepsilon_\partial$ and $\varepsilon_{\rm Gal}$ of~\eqref{eq:diagnostics} on every time slice, which lets us attribute any breakdown to the derivative or the field expansion.

\subsubsection{Gaussian initial-data family}
\label{sec:Gaussian-ID}

Our first family of initial data parametrises an ingoing radial Gaussian pulse, i.e., 
\begin{align}
    \phi(t,r) 
    &= 
    A\,\frac{(r+t)^3}{r}\,e^{-(t + r-r_0)^2/\sigma^2}\;.
    \notag\\
    \label{eq:initial-data}
\end{align}
 Taking the respective spatial and/or time derivatives of this function, and subsecuently evaluating at $t=0$ allows to compute the initial data for the $\psi$ and $A_a$ fields.
 
This choice corresponds to the free ingoing spherical wave condition, which in turn implies $\psi(0,r)=\partial_{t}\psi(0,r)=0$.
This family of initial data is characterised by an amplitude $A$, an initial radius $r_0$, and a width $\sigma$. Since $A$ multiplies $r^2$ in~\eqref{eq:initial-data}, it has dimension $[A]=-3$, and hence $A r_0^3$ is dimensionless. For pulses initially well separated from the origin, the evolution becomes insensitive to $r_0$ at fixed $A r_0^3$, up to the expected free-propagation rescaling. Together with the theory parameter recalled above, the relevant dimensionless ratios are therefore
\begin{align}
    \lbrace\tilde{A},\;\tilde{\sigma},\;\tilde{\beta}\rbrace
    =
    \left\lbrace
        A r_0^3,\;
        \frac{\sigma}{\alpha^{1/3}},\;
        \frac{\beta}{\alpha^{2/3}}
    \right\rbrace
    \;.
    \label{eq:Gaussian-dimensionless-ratios}
\end{align}
In practice, this family of initial data lets us probe the two expansions of~\cref{sec:multi-scale-EFT} independently: Small $\sigma/\alpha^{1/3}$ probes the derivative expansion through high-frequency data, while broad pulses with large $A r_0^3$ probe the field expansion by accumulating large field values as the field approaches the centre of the domain.

For representative fixed $\tilde{\sigma}{=}\sigma/\alpha^{1/3}$ and $\tilde{\beta}{=}\beta/\alpha^{2/3}$, we bisect for the dimensionless critical amplitude $\tilde{A}_\text{crit}(\tilde{\sigma},\tilde{\beta})$ at which the respective evolution first breaks (maintaining at most $1\%$ normalized relative error), that is, where the unregularised evolution ($\beta{=}0$) develops an ill-posed region, or where the regularised evolution ($\beta{\neq}0$) develops unbounded growth from the tachyonic instability of~\cref{sec:regularisation}. To resolve all relevant scales while remaining efficient, we set $r_0{=}4\sigma$, the domain size $L{=}4r_0$, and the number of spatial points to $N{=}64\times L/\min(\sigma,\sqrt{\beta})$.
This makes sure that both the pulse width $\sigma$ and the Compton wavelength of the ghost $\sqrt{\beta}$ are resolved by at least 64 grid points.

\subsubsection{Initial data with a screened source}
\label{sec:screening-ID}

To probe screening, we also investigate an initial-data family in the presence of a prescribed conformally coupled source. 
To be specific, we consider a conformally coupled source Lagrangian $\mathcal{L}_\text{source}{=}\gamma\phi T$, where $T$ is the trace of an energy momentum tensor with mass dimension $[T]{=}4$, and hence the conformal coupling $\gamma$ has mass dimension $[\gamma]{=}-1$. In the context of EFT, this corresponds to $\gamma=c_\gamma/\Lambda_\text{source}$ with $c_\gamma$ denoting the respective dimensionless EFT coefficient and $\Lambda_\text{source}$ the EFT scale at which the source term couples. In general $\Lambda_\text{source}$ may be distinct from $\Lambda_\partial$ and $\Lambda_\phi$. If one considers the Galileon as a proxy for the gravitational interaction, this scale would be interpreted as the Planck scale. In the following, we will see that $\Lambda_\text{source}$ is degenerate with the source amplitude and that they jointly set the effective source strength, which we assume to be independent of the other scales in the theory.
In the spherically symmetric equations, the source enters with an effective source term $J{=}\gamma T$.
For definiteness, we consider the source to be a centred Gaussian profile,
\begin{align}
    T(r)
    =
    - T_0\,e^{-r^2/s_g^2}\;,
    \label{eq:screened-source-profile}
\end{align}
with characteristic amplitude $T_0$ and width $s_g$. The scalar equation contains $\gamma T$ with the same dimension as $\Box\phi$, i.e., $[\Box\phi]{=}3$
as well as $[\gamma T]{=}3$ and hence $[\gamma T_0]{=}3$. The corresponding dimensionless source strength is therefore $\gamma T_0 \alpha$,
up to the overall sign convention in~\cref{eq:screened-source-profile}. Together with the source width $s_g$ and the theory parameter space recalled above, the screened-source family is again characterised by three dimensionless ratios
\begin{align}
    \lbrace\tilde{J},\;\tilde{s}_g,\;\tilde{\beta}\rbrace
    =
    \left\lbrace
        \gamma T_0\alpha,\;
        \frac{s_g}{\alpha^{1/3}},\;
        \frac{\beta}{\alpha^{2/3}}
    \right\rbrace
    \;.
    \label{eq:screened-dimensionless-ratios}
\end{align}
For this family of runs, we initialise the scalar either on the partial vacuum ($\phi(0,r)=0$, $\partial_t\phi(0,r)=0$) or on the screened solution of the unregularised theory. We then evolve in the presence of the fixed external source in~\cref{eq:screened-source-profile}.

We note that the conformal source coupling is not invariant under the perturbative field redefinition in~\cref{eq:field-redef} and in fact generates a dimension-$7$ term, i.e., 
$
    \frac{c_\gamma}{\Lambda_\text{source}}\,\phi\,T
    \rightarrow
    \frac{c_\gamma}{\Lambda_\text{source}}\left(
        \phi
        + \frac{c_\beta}{\Lambda_\partial}\,(\Box\phi)
    \right)\,T
$. However, as long as the field redefinition remains perturbative, the respective transformation of the static source is perturbatively negligible as well. Thus we will neglect such corrections and couple the same static source on both sides of the field redefinition.

Overall, this family of initial data is designed to test whether the partial vacuum can dynamically transition to a screened configuration, and, if so, whether this transition and the resulting screened stationary final state remain within the regime of validity of the EFT.

\section{Results}
\label{sec:results}

%
\begin{figure}
    \begin{centering}
        \includegraphics[width=1\linewidth]{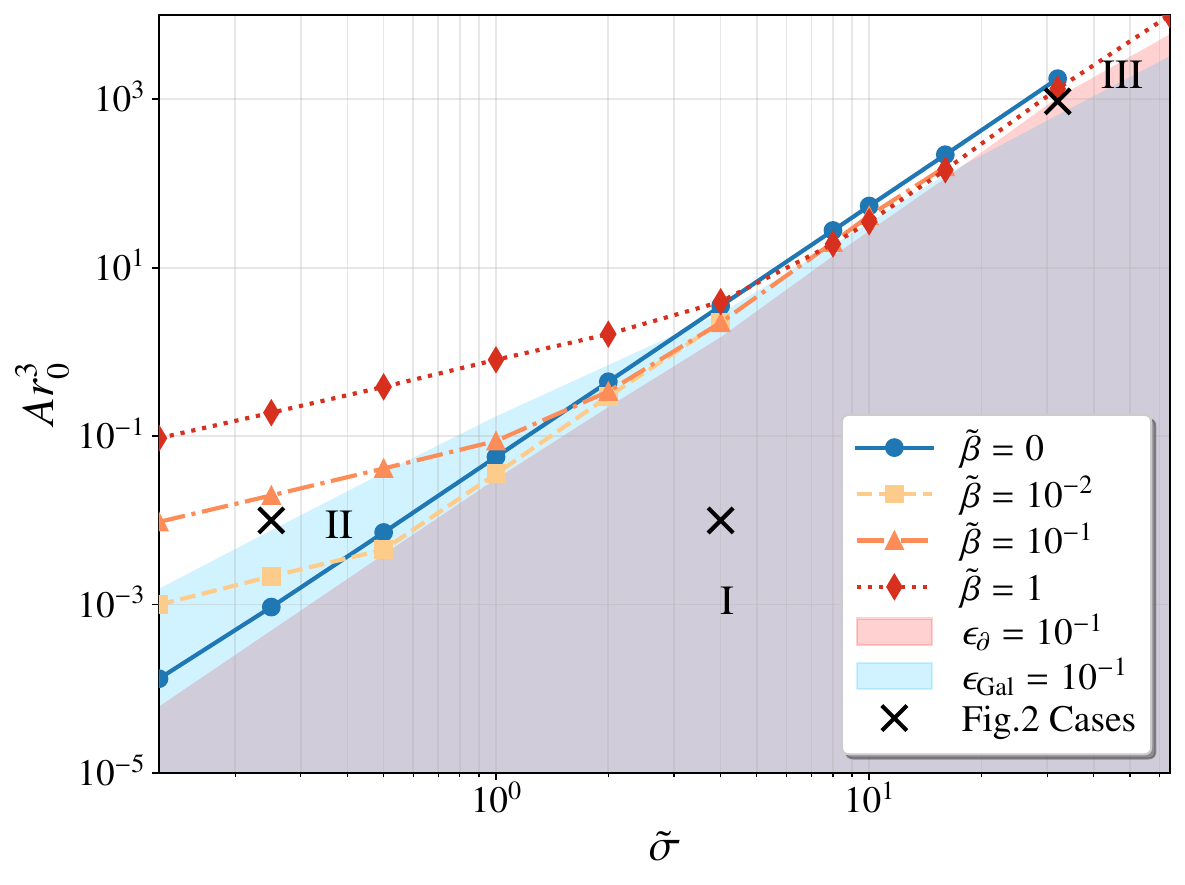}
    \end{centering}
    \vspace*{-2em}
    \caption{\label{fig:EFT-sketch}
    Dimensionless parameter space $(\tilde{\sigma}{\equiv}\sigma/\alpha^{1/3},\tilde{A}{\equiv}Ar_0^3)$ of the Gaussian initial-data family~\eqref{eq:initial-data}. The curves show the critical amplitude $\tilde{A}_\text{crit}(\tilde{\sigma},\tilde{\beta})$ at which the evolution breaks, for different $\tilde{\beta}{\equiv}\beta/\alpha^{2/3}$. The unregularised theory is ($\tilde{\beta}{=}0$) and for the regularised theories we choose $\tilde{\beta}{=}10^{-2},10^{-1},1$. For the reference value $\tilde{\beta}=1$, which corresponds to a single-scale EFT, the shaded bands mark where the derivative and field expansions reach $\varepsilon_\partial{\leqslant}10^{-1}$ (red) and $\varepsilon_{\rm Gal}{\leqslant}10^{-1}$ (blue). At large width (region~III) the critical amplitude is essentially independent of $\tilde{\beta}$, so regularisation does not extend the evolution; at small width (region~II) the contours fan out with $\tilde{\beta}$, so a heavier regularising term pushes the breakdown to larger amplitude. Crosses indicate the cases detailed in~\cref{fig:compare_energies}.
    }
\end{figure}
\begin{figure}
    \centering
    \includegraphics[width=1\linewidth]{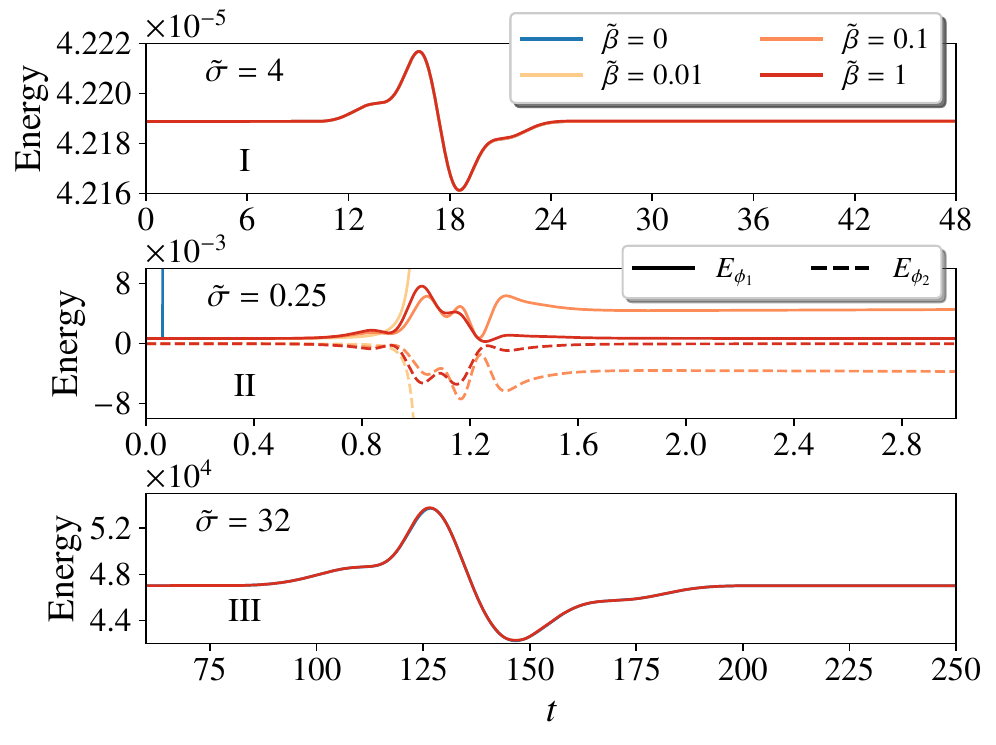}
    \vspace*{-2em}
    \caption{\label{fig:compare_energies}
    Energy components for $\tilde \beta{\equiv}\beta/\alpha^{2/3}{=}0,\,10^{-2},\,10^{-1},\,1$, for exemplary cases in region~I (top panel), region~II (middle panel), and region~III (bottom panel), marked by crosses in~\cref{fig:EFT-sketch}. Solid and dashed curves show the two free components $E_{\phi_1}$ and $E_{\phi_2}$ of~\eqref{eq:energy-budget}. In regimes~I and~III, the curves for all $\tilde \beta$ collapse onto one another, so the formulations are equivalent (in regime~III we only have $\tilde \beta=0,~1$ as the remaining $\tilde \beta$s are numerically costly to resolve for large $\tilde \sigma$); in regime~II they separate, and the ghost component $E_{\phi_2}$ grows, signalling that field-redefinition equivalence has failed.
    }
\end{figure}
In Fig.~\ref{fig:EFT-sketch}, we present, for several values of $\beta/\alpha^{2/3}$, the critical amplitude $A_{\text{crit}}r_0^3$ as a function of the width $\sigma/\alpha^{1/3}$ of the Gaussian pulse~\cref{eq:initial-data}. For $\beta/\alpha^{2/3}{=}1$, the shaded regions mark where the derivative and the field expansions reach $\epsilon_{\partial}{\leq}10^{-1}$ and $\epsilon_{\text{Gal}}{\leq}10^{-1}$, respectively. In the following, we will discuss the results in~\cref{fig:EFT-sketch} from both perspectives developed in~\cref{sec:regularisation} and~\cref{sec:EFT}. Qualitatively similar results also hold for conformally coupled sources and the initial-data family in~\cref{eq:screened-source-profile}. 

\subsection{Regularising Galileons}
\label{sec:results:regularisation}

Here, we discuss the results in Fig.~\ref{fig:EFT-sketch}, from the perspective of~\cref{sec:regularisation}. We observe that for sufficiently small amplitudes, the unregularized theory admits a well-posed evolution, and we do not encounter any runaway tachyonic instability in the regularized theory (see ~\cref{eq:eff-masses}). In this case, it is possible to take the $\beta \to 0$ limit and recover the dynamics of the unregularized theory. We exemplify this behaviour in the top panel of Fig.~\ref{fig:compare_energies}. 
At each timestep, the energy in the $\phi_1$ mode (see~\cref{eq:energy-budget}) converges\footnote{We use the word convergence here without having determined a specific convergence rate and just to state that the solution point wise approaches the $\beta=0$ case. In particular, we have not specifically determined how the convergence of the solution is related to the decoupling of the extra mode.} to that of the unregularised theory as $\beta\to0$. Equivalently, the energy in the $\phi_2$ (ghost) mode converges to zero as $\beta\to0$. In this sense (see~\cref{sec:regularisation}), the regularisation is successful.
In~\cref{fig:EFT-sketch}, this corresponds to region~I, and the EFT interpretation offers a precise definition of this region and an underlying explanation for this behaviour; see~\cref{sec:results:field-redefinition-equivalence}.

At the same time, the successful regularisation does not extend into a regime in which the unregularised theory becomes ill-posed. 
As we increase the amplitude, we find that the unregularised theory develops an ill-posed region, and for large enough $\beta$, we can continue the evolution beyond this point. In fact, for sufficiently large $\beta$, one can apparently continue the evolution for all future time. 
However, the limit $\beta \to 0$ leads to an unquenched tachyonic instability and, hence, to a finite-time divergence. 
This can be seen in~\cref{fig:EFT-sketch}. 
For any value of $\sigma/\alpha^{1/3}$, as one decreases the value of $\beta$, the critical amplitude eventually always drops below the critical amplitude of the unregularised theory.
For $\sigma/\alpha^{1/3}{\leqslant}2$, this occurs already for the simulated $\beta/\alpha^{2/3}{=}10^{-2}$ case. However, we see no reason why this trend should not continue for smaller values of $\sigma/\alpha^{1/3}$ and, respectively, smaller values of $\beta/\alpha^{2/3}$. 

While we have not exhausted the full parameter space of the initial data family in~\cref{eq:screened-dimensionless-ratios}, we have tested $s_g/\alpha^{1/3}{=}32$ and find the same behaviour for the conformally coupled source~\eqref{eq:screened-source-profile}.

\subsection{Field redefinition equivalence}
\label{sec:results:field-redefinition-equivalence}

With the diagnostics and the Gaussian initial-data family in place, we can connect the breakdown of the derivative and field expansions, characterised by $\varepsilon_\partial$ and $\varepsilon_{\rm Gal}$, to the validity of the field redefinition~\eqref{eq:field-redef} and to the respective breakdown of the unregularised and regularised evolutions. 

Three regimes emerge, distinguished by which expansion first leaves its regime of validity. Strictly, the regime boundaries depend on $\beta$ itself. We choose to identify them with respect to $\tilde \beta{=}1$, which corresponds to a single-scale EFT.

\subsubsection{Regime~I: the EFT regime}
\label{subsec:regime1}
For sufficiently small $A$, the nonlinearities stay subdominant, $\varepsilon_{\rm Gal}{\ll}1$ and $\varepsilon_\partial{\ll}1$ throughout the domain, including where the field piles up at the centre. This is the regime of validity of the EFT. Here the unregularised and all regularised evolutions are physically equivalent (top panel of~\cref{fig:compare_energies}), as long as the field redefinition~\eqref{eq:field-redef} stays within the EFT, $|c_\text{redef}\Lambda_\partial^{-2}\Box\phi|{\ll}|\phi|$. This includes the redefinition from the ``natural EFT'' ($c_\alpha{\sim}1$, $c_\beta{\sim}1$) to the unregularised theory. No ill-posed region forms in the unregularised case, and in the regularised case the ghost smoothly decouples as $\beta\to0^+$. Both statements confirm nonlinear field-redefinition equivalence deep inside the EFT.

\subsubsection{Regime~II: probing the high-frequency cutoff}
\label{subsec:regime2}
When the local field values push $\varepsilon_\partial{\sim}1$ while $\varepsilon_{\rm Gal}{\ll}1$, the derivative expansion can no longer be trusted. In the Gaussian family, this happens for small width, $\sigma{\ll}1$, that is, high-frequency initial data (region~II of~\cref{fig:EFT-sketch}). We conclude that the ill-posed regions of the unregularised theory form because a field redefinition has been performed in a local region where the derivative expansion is sensitive to the cutoff. Here, regularisation allows the evolution to continue into a regime where the unregularised evolution breaks down and $\varepsilon_\partial{>}1$. The two formulations are genuinely inequivalent in this regime, and the split is visible in the energy budget: the middle panel of~\cref{fig:compare_energies} shows the free components separating for different values of $\beta$, with the ghost component $E_{\phi_2}$ no longer negligible.

\subsubsection{Regime~III: probing the strong-field cutoff}
\label{subsec:regime3}
When instead the local field values push $\varepsilon_{\rm Gal}{\sim}1$ while $\varepsilon_\partial{\ll}1$, the field expansion can no longer be trusted. In the Gaussian family, this happens for large width, $\sigma{\gg}1$ (region~III of~\cref{fig:EFT-sketch}). Our numerics indicate that field-redefinition equivalence is maintained even here, in the strong-field regime, as long as the derivative expansion stays valid. This is consistent with the field redefinition~\eqref{eq:field-redef} being linear in the field but nonlinear in derivatives: it is controlled by $\varepsilon_\partial$, not by $\varepsilon_{\rm Gal}$. In agreement with this, the regularising term does not let us continue the evolution where the unregularised evolution breaks, unless $\tilde\beta \gg 1$, that is, the ghost is made very light.\footnote{A sufficiently light ghost can always avoid the instability, at the cost of propagating a mode well within the EFT window.} The bottom panel of~\cref{fig:compare_energies} confirms that for $\beta=0$ and $\beta=1$ (the remaining values of $\beta$ are numerically costly to resolve for large $\sigma$) they track the same evolution. Whether further interactions at higher orders in the field expansion can quench this dynamical instability and thereby delay the breakdown will be addressed in separate work.

We conclude that field-redefinition equivalence holds whenever the derivative expansion remains valid, and even in the strong-field regime, at least within the Gaussian family of initial data.

\subsection{Screening}
\label{sec:screening}

\begin{figure}
    \centering
    \includegraphics[width=1\linewidth]{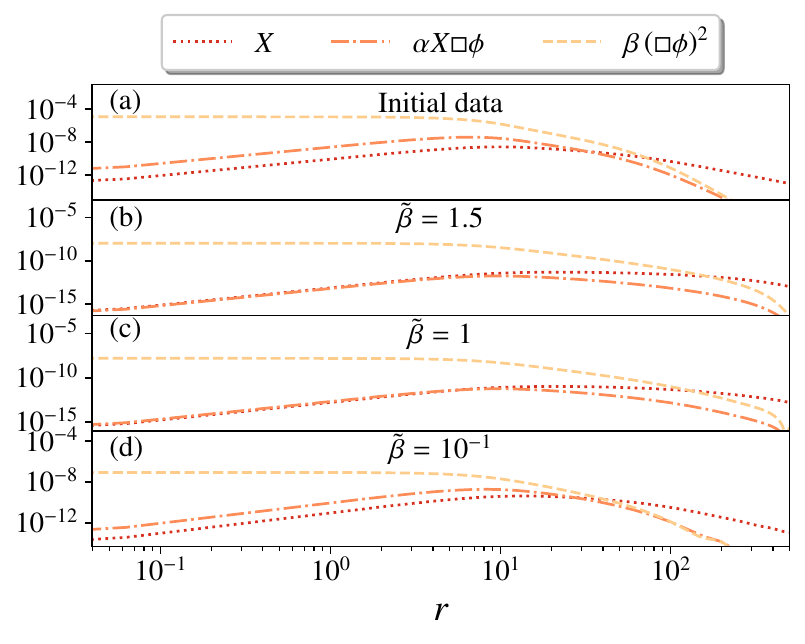}
    \vspace*{-2.5em}
    \caption{\label{fig:screening}
    Screened solutions represented by the local (in space) values of the kinetic term $X{=}-\tfrac{1}{2}(\partial\phi)^2$, the Galileon term $\alpha X\Box\phi$, and the ghost term $\beta(\Box\phi)^2$, for different $\beta$s, with $\tilde\beta \equiv \beta / \alpha^{2/3}$.
    Panel~(a) shows the static screened solution of the unregularised ($\beta{=}0$) theory; for the ghost term, we evaluate with $\tilde\beta{=}1$. Panels~(b) and~(c) show the final states of transitions from the initial data in~(a) to the true screened solution of the respective regularised theory. Panel~(d) shows the final state of an example in which dynamical relaxation is needed to avoid an intermediate unquenched tachyon. In every panel, the ghost term (dashed) sits above both the Galileon and the kinetic term throughout the screened region.
    }
\end{figure}
Galileon interactions are a standard model for screening. In the usual setup, the ghost term is not considered, and it has been argued~\cite{Nicolis:2004qq,Luty:2003vm,Joyce:2014kja} that one may expect it to remain consistently small in the vicinity of the screened solution. 
However, in our simulations we find that no such regime exists and that, as one might have anticipated from the EFT scaling below~\cref{eq:Lagrangian-EFT-Galileon}, the ghost term is dominant throughout the screened region; see~\cref{fig:screening}. 

The spherically symmetric setup of~\cref{sec:screening-ID} allows us to probe this expectation directly by working with the full nonlinear solution rather than with perturbative estimates. 
To source a screened configuration, we pick a sufficiently large dimensionless source amplitude $(\gamma T_0\alpha)$ and sufficiently small dimensionless source width $s_g/\alpha^{1/3}$. To be specific, we chose $(\gamma T_0\alpha){=}10^6$ and $s_g/\alpha^{1/3}{=}32$ and constructed screened solutions of the unregularised Galileon theory~\cite{Nicolis:2004qq,Brito:2014ifa}.
The top panel in~\cref{fig:screening} shows the three interaction terms of~\cref{eq:Lagrangian-EFT-Galileon} on this solution. At large radii, the kinetic term dominates, and the radial profile falls off as in the free theory. At small radii, we recover the screened region, i.e., a central region in which the Galileon term $\alpha X\Box\phi$ dominates over the kinetic term $X$. Throughout this region, however, the ghost term $\beta(\Box\phi)^2$ is larger still, see panel~(a) of~\cref{fig:screening}, at least if $\beta/\alpha^{2/3}\sim1$ as expected in EFT. We must therefore conclude that the actual nonlinear solution does not support the expectation that the ghost is negligible.

Interestingly, we can still obtain consistent screened solutions in the regularised theory. We generate them by choosing a suitable matter profile and evolving the regularised system. We initiate the evolution either from the screened solution of the unregularised theory\footnote{Using this solution and the assumption that time derivatives vanish, we can then compute the initial data for the $\psi$ and $A_{a}$ fields.} or from the EFT vacuum, neither of which is an equilibrium of the regularised theory. As expected, we observe a transient in which the local field settles to the true solution by radiating a nonlinear pulse out towards spatial infinity. In both cases we reach the true screened solution of the regularised theory, provided the intermediate transient is not too violent; see panels~(b) and~(c) of~\cref{fig:screening}.

If instead the initial profile, be it the trivial vacuum or the unregularised screened solution, is too far from the true regularised solution, an unquenched tachyon obstructs the transition. Even then, we can reach the true solution by dynamical relaxation. For a large enough $\beta$, we start from a screened solution of the cubic Galileon theory ($\beta{=}0$) and evolve until we reach the new equilibrium of the theory with $\beta{\neq}0$. Starting from this equilibrium, we then use a smooth function of time\footnote{We use $  \beta(t\le t_{i})=\beta_{i}$, $ \beta(t_i<t<t_f)=\beta_{i}(1-s + s F)$ and $\beta(t\ge t_f)=\beta_i F$, where, $s=\frac{1}{2}\left(1 - \cos \left( \pi \frac{t -t_i}{t_f-t_{i}}\right)\right)$, $F<1$, $\beta_i$ is the initial value of $\beta$ and $t_i$ and $t_f$ determine the initial and final time of the transition.} to reduce $\beta$ dynamically, and take the final configuration as initial data, see panel~(d) of~\cref{fig:screening}. Once the evolution has settled around the true screened solution of the regularised theory, the solution appears to be stable under small perturbations.

\section{Conclusions}
\label{sec:conclusions}

We have solved the spherically symmetric dynamics of a scalar with Galileon symmetry to demonstrate how regularisation operates and how it can be interpreted within the context of an EFT. 
This theory is particularly tractable since it allows us to cleanly distinguish between a single cubic interaction term, $(\partial\phi)^2\Box\phi$, and a single regularisation term, $(\Box\phi)^2$. Further, these two terms correspond to the leading terms in the respective EFT and distinguish between expansions in derivatives and in the fields. 
The regularisation term can be introduced, removed, or altered by perturbative linear field redefinitions.

For the unregularised theory in $(3{+}1)$ dimensions, we cannot rely on a generic local well-posedness result for all sufficiently smooth initial data. In practice, however, we find convergent evolution whenever the initial data remains within the EFT's regime of validity. 

At least when restricting to spherical symmetry, this can be understood precisely, since the respective characteristic speeds can be obtained analytically (see~\cref{eq:characteristic-speeds}) and thus provide a direct local diagnostic. Real, finite, and non-degenerate characteristic speeds imply strong hyperbolicity and thus guarantee local well-posedness, whereas divergent characteristic speeds imply a loss of hyperbolicity and thus local ill-posedness.

In actual numerical evolution, there are two potential sources of such ill-posedness. First, the physical solution may correspond to ill-posed initial data or dynamically evolve from a well-posed to an ill-posed configuration. Second, numerical noise may generate random fluctuations that, if significant enough, can lead to ill-posed characteristic speeds. We highlight that this second source of ill-posedness does not obstruct the evolution as long as the numerical noise fluctuations remain small enough not to produce divergent characteristic speeds. Based on~\cref{eq:characteristic-speeds}, we estimate this condition to hold whenever
\begin{align}
    \alpha
    \lesssim
    \frac{\text{resolution}^2}{\text{machine precision}}.
\end{align}
Provided that both the physical evolution and the numerical fluctuations avoid configurations with divergent characteristic speeds, the unregularised theory therefore behaves as an ``effectively well-posed'' system, despite the absence of a generic local well-posedness statement.

Regularisation introduces a fiducial parameter $\beta$ and an associated fiducial massive mode $\psi$ with mass $m_\psi\sim1/\sqrt{\beta}$, such that one can rewrite the system into a set of nonlinear wave equations and guarantee well-posedness for all sufficiently smooth initial data.
We have explicitly demonstrated this high-frequency regularisation for a Gaussian family of initial data (see~\cref{fig:EFT-sketch}), where one can see that any finite $\beta$ improves the high-frequency/small-width limit. 
In general, we call regularisation successful if the solution converges for sufficiently small $\beta$. The physical mechanism behind this convergence is the dynamical decoupling of the associated fiducial mode.
For the Galileon theory, we have confirmed explicitly that whenever regularisation is successful, it recovers the unregularised solution. Conversely, whenever the unregularised theory becomes ill-posed, a tachyon develops for sufficiently small $\beta$, obstructing successful regularisation.

At least for the Galileon example at hand, the EFT provides an underlying explanation for the occurrence of ill-posedness. More specifically, we find that ill-posedness occurs if and only if the derivative expansion breaks.
EFT decoupling and field-redefinition equivalence seem to explain the regime in which regularisation succeeds. In particular, we confirm that whenever $\beta{<}\Lambda_\partial^2$, the fiducial degree of freedom decouples, and regularisation is successful. Equivalently, we confirm field-redefinition equivalence whenever the respective linear field redefinition remains within the validity of the derivative expansion.

A related example with a known UV completion has been addressed in~\cite{Figueras:2025wtx}. While not formally demonstrated by the authors, we expect that field-redefinition equivalence holds there as well, as long as one remains within the regime of validity of the derivative expansion. As a result, we expect that even the unregularised theory truncation remains ``effectively well-posed'' and can be evolved on suitable initial data and with sufficiently high machine precision.

It remains an open question how far these flat-space scalar field results extend beyond flat space and to more general EFTs, in particular, those involving a dynamical metric. 
For instance, it would be fruitful to repeat the present analysis for the scalar-tensor EFTs mentioned in the introduction.

We have also added a conformally coupled source term to the Galileon EFT to explore screening. Static screened solutions for a point mass source have been obtained in~\cite{Nicolis:2004qq,Brito:2014ifa}. Further, it has been argued~\cite{Nicolis:2004qq,Luty:2003vm,Joyce:2014kja} that these solutions are within the EFT and that the ghost term can be consistently neglected on the screened solution. In our spherically symmetric setup, we can directly explore this expectation by solving the nonlinear dynamics by including the conformally coupled matter source (see~\cref{sec:screening-ID}).

We find that, for any non-vanishing $\beta$, the ghost term always dominates in the core of the screened solution (see~\cref{fig:screening}). Under the additional assumption of naturalness, i.e., for $\beta/\alpha^{2/3}\sim 1$, the ghost dominates not just within the core but throughout the entire screening region.
This holds both if we evaluate the respective terms on the screened solutions of the unregularised theory and if we evaluate them on the screened solutions of the regularised theory.
We conclude that the ghost is excited and cannot be neglected unless one considers unnaturally small $\beta/\alpha^{2/3}\ll 1$ and neglects the region in the core.
In the context of EFT, we see no generic justification for $\beta/\alpha^{2/3}\ll 1$ and thus conclude that the screened solutions do not belong to the standard Galileon EFT but rather to an EFT in which the ghost is an active low-energy degree of freedom.

The presence of the ghost also explains why the screened solutions cannot be reached dynamically. More specifically, we find that starting from the partial vacuum ($\phi=0$, $\partial_t\phi=0$), one cannot dynamically reach a screened solution in the unregularised theory ($\beta=0$) because an ill-posed region obstructs the transition. This is consistent with the conclusions we drew from the Gaussian initial data: the required field redefinition that removes the ghost cannot remain within the regime of validity of the derivative expansion, since the ghost is excited on the screened solution. This explains why the respective unregularised evolution encounters an ill-posed region.

It remains to be explored how far similar consistency issues affect nonlinear phenomena in EFTs more generally.
Specifically, it seems prudent to verify whether all higher-derivative corrections that naturally arise in an EFT can be neglected around the respective screened solutions and other proposed nonlinear phenomena.

Finally, we emphasize that both the regularised and unregularised dynamics eventually break down, however in very different ways. In the ghost-free (unregularised) formulation, sufficiently high-frequency data, even at arbitrarily small amplitude, leads to ill-posedness. In the ghostly (regularised) formulation, sufficiently large amplitudes trigger dynamical instabilities that, if left unquenched, lead to divergences. We regard it as an interesting open avenue to identify interactions that quench these dynamical instabilities; see~\cite{Deffayet:2025lnj,Held:2025fii} for a mechanism based on non-derivative interactions.
\begin{acknowledgments}
We thank \'Aron Kov\'acs and Jan Ko{\.z}uszek for discussions and feedback on an earlier version of this manuscript. FT is supported by the INFN Postdoctoral research agreement No.~27076. R.C. acknowledges support from the European Union’s
Horizon ERC Synergy Grant ``Making Sense of the Un-
expected in the Gravitational-Wave Sky'' (Grant No.
GWSky–101167314)and from the PRIN
2022 grant ``GUVIRP - Gravity tests in the UltraVi-
olet and InfraRed with Pulsar timing''. 
T.S. acknowledges partial support from the STFC Consolidated Grant no. ST/V005596/1 and partial support from the STFC Consolidated Grants no. ST/X000672/1 and UKRI2492.
\end{acknowledgments}

\bibliography{biblio}
\end{document}